\documentclass[11pt,a4paper]{article}

\usepackage[utf8]{inputenc}
\usepackage[T1]{fontenc}
\usepackage{amsmath,amssymb}
\usepackage{graphicx}
\usepackage{booktabs}
\usepackage{tabularx}
\usepackage{url}
\usepackage{hyperref}
\usepackage{tikz}
\usetikzlibrary{arrows.meta, positioning}
\usepackage{microtype}
\usepackage[numbers]{natbib}
\usepackage{algorithm}
\usepackage{algpseudocode}
\usepackage{booktabs}
\usepackage{float}
\usepackage{placeins}

\usepackage[margin=2.5cm, left=3.5cm]{geometry}
\usepackage{lineno}
\usepackage{setspace}
\usepackage{fancyhdr}

\providecommand{\doi}[1]{doi:~\url{#1}}

\hypersetup{
    colorlinks=true,
    linkcolor=blue,
    citecolor=blue,
    urlcolor=blue
}

\title{\textbf{A precision-driven evaluation protocol for author name disambiguation\\[6pt]
{\large Evaluating author disambiguation with focus on lumping and splitting errors}}
}

\author{
Francesca Cappelli$^{1}$ \quad Giovanni Colavizza$^{1,2,3}$ \quad Silvio Peroni$^{1,3,4}$
}

\date{%
$^1$Department of Classical Philology and Italian Studies, University of Bologna, Bologna, Italy\\[4pt]
$^2$Department of Communication, University of Copenhagen, Copenhagen, Denmark\\[4pt]
$^3$Digital Humanities Advanced Research Center (DHARC), University of Bologna, Bologna, Italy\\[4pt]
$^4$Research Centre for Open Scholarly Metadata, Department of Classical Philology and Italian Studies, University of Bologna, Bologna, Italy\\[4pt]
}

\begin{document}

\maketitle

\begin{abstract}
Author Name Disambiguation systems are commonly evaluated using pairwise and B$^3$ metrics, which summarize clustering quality through aggregate precision and recall but do not explicitly distinguish between lumping--merging publications from different authors--and splitting--fragmenting the publications of a single author. Because these errors can have asymmetric consequences in bibliographic applications, we propose a precision-driven evaluation framework that decomposes clustering quality into complementary instance-level and cluster-level signals. The framework introduces cluster-level Lumping Error and Splitting Error to separately quantify contamination and fragmentation, combines them with standard pairwise and B$^3$ metrics through a Composite Score, and uses the resulting measures for model selection. We characterize the diagnostic behavior of the framework through a controlled synthetic case study and assess the sensitivity of the Composite Score to its weighting scheme. The results show that the proposed structural measures distinguish error patterns that aggregate metrics can mask, while the sensitivity analysis identifies a stable region for the adopted weighting configuration. The framework is released as an open-source Python pipeline.
\end{abstract}

\section{Introduction}
 
Author Name Disambiguation (AND) is the task of partitioning a set of publications sharing an ambiguous author name into disjoint clusters, each corresponding to a distinct real-world individual \cite{ferreira2012brief, rodrigues2024author}. It is a fundamental preprocessing step in bibliographic data management, affecting the quality of scholarly profiles, citation analysis, and research assessment systems \cite{strotmann2012author, abramo2023how}. In large-scale open infrastructures such as OpenCitations \cite{peroni2020opencitations}, errors in AND propagate silently across downstream analyses, making robust and interpretable evaluation essential. With the rapid evolution of deep learning and representation learning architectures applied to AND \cite{cappelli2025deep}, establishing a reliable, structural evaluation protocol has become fundamental\cite{subramanian2021s2and, chen2023web}.
 
Despite its importance, AND evaluation lacks a shared, single standardized framework consistently adopted across studies\cite{kim2021}. Most systems are evaluated using canonical pairwise metrics, which operate by enumerating all pairs of publications and checking whether they are correctly co-clustered. While intuitive, pairwise metrics have well-known structural limitations \cite{kim2019}: they are quadratically sensitive to cluster size, disproportionately emphasizing authors with large publication sets; they tend to reward degenerate solutions that produce many singleton clusters; and crucially, they treat a missed merge and an incorrect merge as equivalent \cite{kim2018evaluating}.

$B^{3}$ metrics \cite{bagga1998entity}, adopted in AND-specific contexts \cite{levin2012, louppe2016, kim2019}, were introduced to address the sensitivity to cluster size imbalance. Unlike pairwise metrics, $B^{3}$ operates at the instance level, computing for each publication the fraction of its predicted cluster that overlaps with its ground-truth cluster, and vice versa. This formulation satisfies stronger formal desiderata--homogeneity, completeness, and robustness to cluster size imbalance \cite{amigo2009comparison}--making it a more reliable signal in settings where author cluster sizes are highly skewed. Yet $B^{3}$ shares with pairwise metrics a fundamental limitation: being an aggregate measure, it quantifies how much of each instance's cluster membership is correct, but does not characterize how clustering fails structurally when errors occur. The same holds for the K-metric, another widely used aggregate measure, which has been shown to yield precision and recall identical to $B^{3}$, differing only in how the two are combined into a single score.

This shared limitation becomes particularly consequential when the two qualitatively different failure modes of clustering are considered \cite{torvik2009}. Lumping occurs when publications from distinct authors are incorrectly merged into the same cluster, contaminating author profiles with foreign attributions. Splitting occurs when a single author's publications are fragmented across multiple clusters, reducing the completeness of their profile. In most bibliographic applications, these two error types are asymmetric in their consequences: lumping introduces false positive associations that are difficult to correct downstream, whereas splitting is typically more recoverable. In production systems managing large-scale disambiguation \cite{zhang2018}, this precision-first stance has proven essential. Therefore, this asymmetry motivates a precision-first evaluation stance, in which cluster purity is prioritized over recall.
 
In this paper, we present a structured evaluation framework for AND that addresses these limitations. The framework has been devised and refined through an experimental phase in which multiple embedding models and graph-based configurations were evaluated on two AND benchmarks with contrasting metadata availability. It is presented here independently, as a reusable protocol, so that it can be adopted beyond the specific experimental context in which it was developed. It is composed of four components: (1) standard pairwise and $B^{3}$ metrics, retained for comparability with prior work; (2) cluster-level reformulations of lumping and splitting errors; (3) a Composite Score that integrates all components into a single ranking criterion for model selection; and (4) an empirical characterization of the framework's own behavior, through a controlled synthetic case study and a sensitivity analysis of the Composite Score's weighting scheme. We present the framework in step-by-step form, discuss its formal properties, characterize its behavior empirically, describe its open-source implementation, and analyze its known limitations and workarounds.

\section{Materials and Methods}

The framework takes as input two cluster assignments over the same set of papers: the predicted clusters $\mathcal{C}_P$, produced by an AND system, and the ground-truth clusters $\mathcal{C}_T$, derived from verified author identifiers. It outputs a structured set of metrics characterizing clustering quality from three complementary perspectives: pairwise consistency, B$^3$ instance-level correctness, and structural error decomposition.

\begin{figure*}[htbp]
\centering
\begin{tikzpicture}[
    node distance=0.8cm and 1.2cm,
    box/.style={draw, rounded corners, align=center,
                minimum width=2.2cm, minimum height=0.85cm, font=\small},
    arrow/.style={-Latex, thick}
]
\node[box] (pred)  {Predicted\\Clusters $\mathcal{C}_P$};
\node[box, right=1.5cm of pred]  (eval)  {\textbf{Unified Evaluation}\\\textbf{Pipeline}};
\node[box, right=1.5cm of eval] (truth) {Ground Truth\\Clusters $\mathcal{C}_T$};

\node[box, below left=1.0cm and 0.2cm of eval] (pair)   {Pairwise Metrics\\TP / FP / FN};
\node[box, left=0.6cm of pair]  (b3)     {B$^3$ Metrics\\Instance-level};
\node[box, right=0.6cm of pair] (struct) {Structural Metrics\\LE / SE};

\node[box, below=1.0cm of pair, minimum width=4.5cm] (score)
    {\textbf{Composite Score}\\$0.3\,F1_{PW} + 0.5\,F1_{B^3} + 0.2\,S_{\text{struct}}$};

\draw[arrow] (pred)   -- (eval);
\draw[arrow] (truth)  -- (eval);
\draw[arrow] (eval)   -- (pair);
\draw[arrow] (eval)   -- (b3);
\draw[arrow] (eval)   -- (struct);
\draw[arrow] (pair)   -- (score);
\draw[arrow] (b3)     |- (score);
\draw[arrow] (struct) |- (score);
\end{tikzpicture}
\caption{Overview of the evaluation pipeline. Predicted and ground-truth cluster assignments feed a unified implementation that computes pairwise, B$^3$, and structural metrics, aggregated into a final Composite Score.}
\label{fig:pipeline}
\end{figure*}
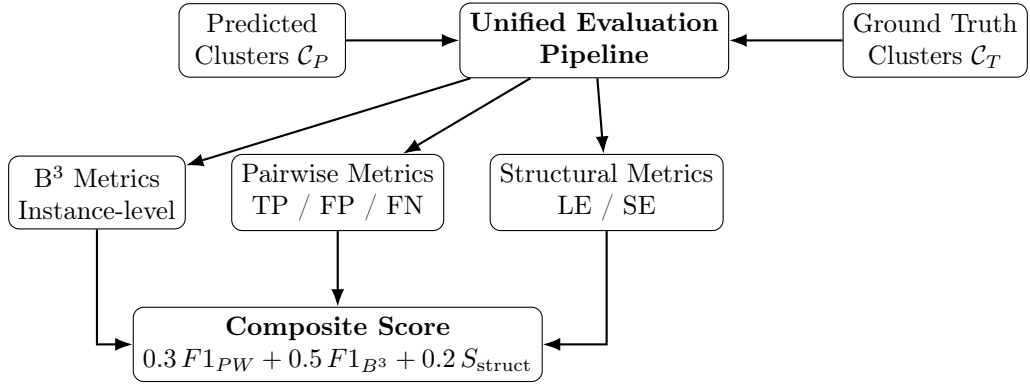

An overview of the framework components is provided in Table~\ref{tab:overview} and Figure~\ref{fig:pipeline}.

\begin{table}[htbp]
\centering
\caption{Summary of the evaluation framework components.}
\label{tab:overview}
\small
\setlength{\tabcolsep}{4pt}
\begin{tabularx}{\columnwidth}{@{}llX@{}}
\toprule
\textbf{Component} & \textbf{Level} & \textbf{Purpose} \\
\midrule
Pairwise F1 & Pair & Global consistency check across all paper pairs \\
B$^3$ F1 & Instance & Cluster purity and completeness per paper \\
Lumping Error & Cluster & Fraction of predicted clusters mixing distinct authors \\
Splitting Error & Cluster & Fraction of ground-truth clusters fragmented across predictions \\
$S_{\text{struct}}$ & Cluster & Aggregate structural quality score \\
Composite Score & All & Single ranking criterion for model selection \\
\bottomrule
\end{tabularx}
\end{table}

Let $N$ be the total number of papers. For each paper $i$, we denote:
\begin{itemize}
    \item $P(i)$: the predicted cluster containing $i$,
    \item $T(i)$: the ground-truth cluster containing $i$.
\end{itemize}

\subsection{Step 1 -- Precision-First Evaluation Principle}
\label{sec:precision-first}

We adopt a precision-first evaluation principle, motivated by the asymmetric impact of clustering errors in AND. Merging publications belonging to different authors (lumping) is typically more detrimental than splitting a single author into multiple clusters, as it directly contaminates author profiles with incorrect attributions.

This assumption is consistent with prior work in entity resolution and clustering, where preserving cluster purity is often prioritized over completeness \cite{saeedi2020}. This principle aligns with AND density-based clustering approaches \cite{ester1996} which emphasize local coherence and purity within clusters. In applications where false associations are difficult to correct downstream, such assumptions are particularly consequential. Therefore, while both precision and recall are reported, particular emphasis is placed on precision and on controlling lumping errors.

\paragraph{Pairwise Precision and Recall.} Let $TP$ be the number of paper pairs correctly placed in the same cluster, $FP$ the number incorrectly placed in the same cluster, and $FN$ the number belonging to the same ground-truth cluster but placed in different predicted clusters:

\begin{equation}
\label{eq:pairwise}
PW_P = \frac{TP}{TP+FP}, \qquad PW_R = \frac{TP}{TP+FN}
\end{equation}

\paragraph{B$^3$ Precision and Recall.}

\begin{equation}
\label{eq:b3p}
B^3_P = \frac{1}{N}\sum_{i=1}^{N} \frac{|P(i)\cap T(i)|}{|P(i)|}
\end{equation}
\begin{equation}
\label{eq:b3r}
B^3_R = \frac{1}{N}\sum_{i=1}^{N} \frac{|P(i)\cap T(i)|}{|T(i)|}
\end{equation}

$B^3_P$ evaluates cluster purity at the instance level and is less sensitive to cluster size imbalance than pairwise precision \cite{amigo2009comparison}.

\paragraph{F1 scores.} For both families, precision and recall are combined into a single score through their harmonic mean:

\begin{equation}
\label{eq:f1}
F1_{X} = \frac{2\, X_P\, X_R}{X_P + X_R}, \qquad X \in \{PW, B^3\}
\end{equation}

\begin{figure*}[htbp]
\centering
\includegraphics[width=\textwidth]{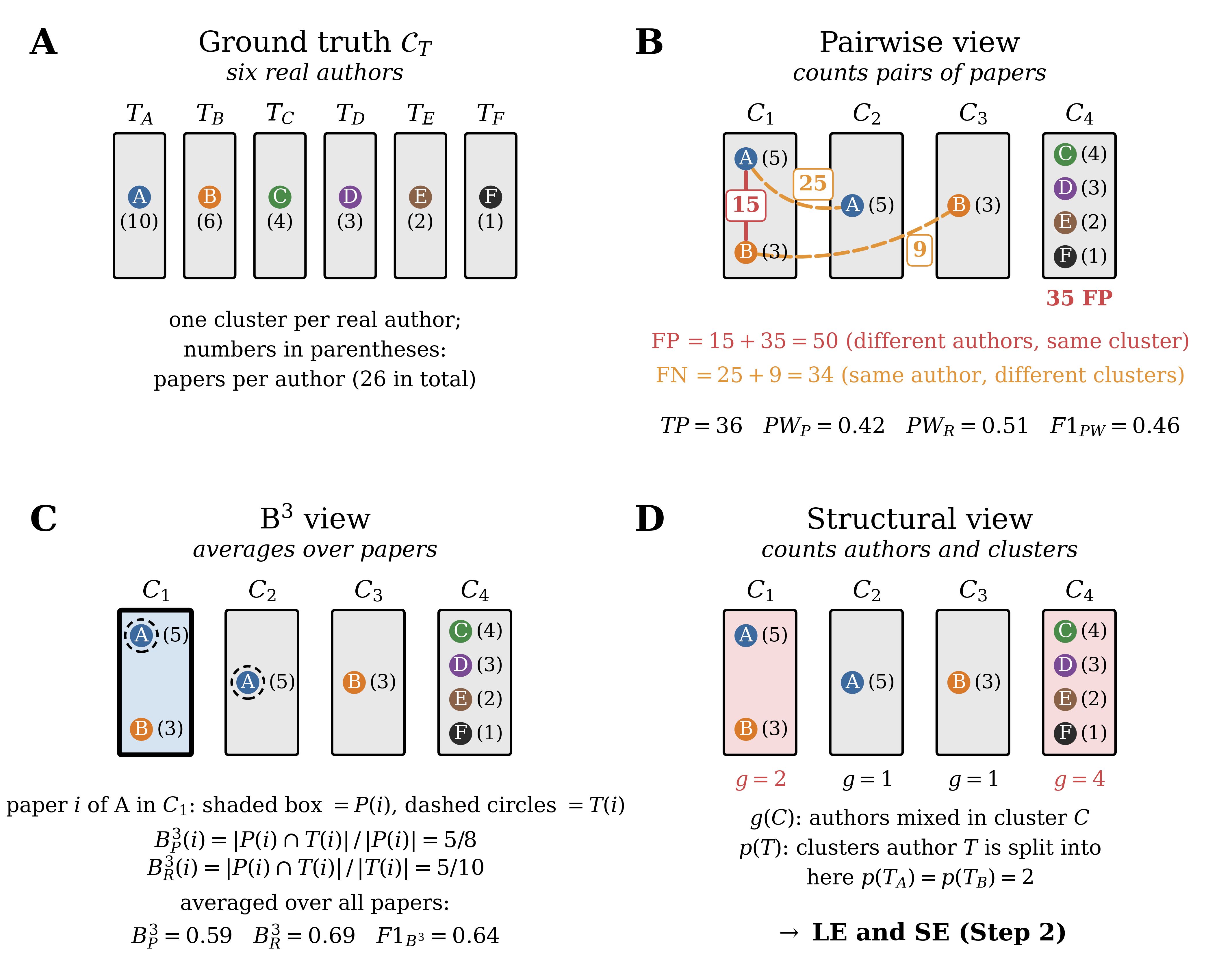} 
\caption{Evaluation perspectives on one ambiguous name. (A) Ground truth $\mathcal{C}_T$: six real authors; numbers in parentheses indicate papers. (B--D) The same predicted clustering $\mathcal{C}_P$ (Case~3 in Figure~\ref{fig:toy2}) seen from three perspectives. (B) Pairwise view: solid red lines mark false positive pairs, dashed orange lines false negative pairs. (C) B$^3$ view: precision and recall for one paper $i$ of A in $C_1$, then averaged over all papers. (D) Structural view: number of distinct authors $g(C)$ in each predicted cluster.}
\label{fig:toy1}
\end{figure*}

\paragraph{Example: Structural Interpretation of Pairwise and B$^3$ Metrics.} To provide an intuitive visual interpretation of traditional evaluation paradigms, Figure~\ref{fig:toy1} considers one ambiguous name shared by six real authors, whose ground-truth clusters $T_A, \dots, T_F$ contain respective number of papers (panel A). An AND system produces four predicted clusters $C_1, \dots, C_4$ that both merge and fragment authors, evaluated from three perspectives (panels B--D).

The example highlights how the two classical families in Eq.~\eqref{eq:pairwise}--\eqref{eq:b3r} extract errors from the same clustering. The pairwise paradigm (Figure~\ref{fig:toy1}B) counts pairs of papers: merging the 5 papers of A with the 3 papers of B in $C_1$ produces $5 \times 3 = 15$ false positive pairs, while splitting A across $C_1$ and $C_2$ produces $5 \times 5 = 25$ false negative pairs. Since the number of pairs grows quadratically with cluster size ($O(N^2)$), errors involving large authors dominate $F1_{PW}$. The B$^3$ paradigm (Figure~\ref{fig:toy1}C) instead takes the point of view of each paper $i$, comparing its predicted cluster $P(i)$ with its ground-truth cluster $T(i)$, and averages over all papers -- so $F1_{B^3}$ tracks how well each paper is placed rather than how many pairs are affected.

While both formulations compress overall clustering quality into a single global scalar, they intrinsically merge distinct structural phenomena. By collapsing precision and recall into a single harmonic mean (F1), both paradigms become blind to the direction of errors: they report a generic drop in accuracy but mask whether the system is failing due to identity contamination or cluster fragmentation. Figure~\ref{fig:toy1}D anticipates the alternative adopted in Section~\ref{sec:structural}: the same clustering is read at the level of clusters and authors, counting how many authors each predicted cluster mixes and how many clusters each author is split into, independently of paper counts.

\subsection{Step 2 -- Structural Error Decomposition}
\label{sec:structural}

Although B$^3$ captures instance-level correctness, it does not explicitly describe the structural nature of clustering errors. Two models may achieve similar B$^3$ scores while exhibiting fundamentally different behaviors: one may over-merge distinct authors into large clusters, while another may over-fragment a single author across many small clusters. To improve interpretability, we decompose errors into two complementary structural components.

Splitting and lumping errors have been studyied and discussed in AND evaluation context as standard diagnostic measures. In the conventional formulation, they are defined at the instance level: lumping counts the name instances incorrectly merged into the same predicted cluster, and splitting counts those separated across multiple clusters.

In practice, instance-level formulations of Lumping Error (LE) and Splitting Error (SE) tend to correlate strongly with B$^3$ precision and recall, limiting their diagnostic value as independent structural signals.

We propose a cluster-level reformulation. Rather than counting misassigned instances, we measure how many distinct ground-truth or predicted clusters are incorrectly mixed or fragmented within each cluster, normalized by the total number of clusters involved. This formulation operates at the topology of the clustering, making it conceptually independent from B$^3$.

Let $\mathcal{C}_P$ be the set of predicted clusters and $\mathcal{C}_T$ the set of ground-truth clusters. For a predicted cluster $C \in \mathcal{C}_P$, let $g(C)$ denote the number of distinct ground-truth clusters represented in $C$. For a ground-truth cluster $T \in \mathcal{C}_T$, let $p(T)$ denote the number of predicted clusters intersecting $T$.

\paragraph{Normalized Lumping Error (LE).} A predicted cluster $C$ is perfectly pure when $g(C)=1$. The lumping error is the fraction of excess ground-truth clusters relative to the total, averaged over all predicted clusters:

\begin{equation}
\label{eq:le}
LE = \frac{1}{|\mathcal{C}_P|}\sum_{C\in\mathcal{C}_P} \frac{\max(0,\, g(C)-1)}{g(C)}
\end{equation}

$LE=0$ when every predicted cluster is pure; $LE\to 1$ when predicted clusters systematically merge many distinct authors.

\paragraph{Normalized Splitting Error (SE).} A ground-truth cluster $T$ is perfectly recovered when $p(T)=1$. The splitting error is the fraction of excess predicted clusters relative to the total, averaged over all ground-truth clusters:

\begin{equation}
\label{eq:se}
SE = \frac{1}{|\mathcal{C}_T|}\sum_{T\in\mathcal{C}_T} \frac{\max(0,\, p(T)-1)}{p(T)}
\end{equation}

$SE=0$ when every ground-truth cluster is fully recovered; $SE\to 1$ when ground-truth clusters are systematically fragmented.

\paragraph{Structural Score.} Both $LE$ and $SE$ lie in $[0,1]$ by construction. We aggregate them into:

\begin{equation}
\label{eq:sstruct}
S_{\text{struct}} = 1 - \frac{LE+SE}{2}
\end{equation}

$S_{\text{struct}}\in[0,1]$, where 1 is a structurally perfect clustering and 0 is a maximally degenerate one.

Crucially, $S_{\text{struct}}$ is conceptually independent from B$^3$: while B$^3$ measures \emph{how much} of each instance's membership is correct, $S_{\text{struct}}$ measures \emph{how many} distinct authors or fragments are incorrectly mixed or separated at the cluster level. A model assigning all publications to a single cluster achieves $B^3_R=1$ but $LE\to1$ as $|\mathcal{C}_T|$ grows. Conversely, a model producing all singletons achieves $B^3_P=1$ but $SE\to1$.

We note already here -- and quantify in Appendix~\ref{sec:sensitivity} -- that Eq.~\eqref{eq:sstruct} weights $LE$ and $SE$ symmetrically. This is a deliberate simplification for interpretability, but it means $S_{\text{struct}}$ does not itself carry the precision-first asymmetry that motivates the rest of the framework (Section~\ref{sec:precision-first}); a configuration with zero lumping but moderate splitting can score below one with moderate lumping and zero splitting on $S_{\text{struct}}$ alone, even though the former is preferable under our stated cost model. We return to the practical consequences of this in Appendix~\ref{sec:sensitivity}.

\begin{figure*}[htbp]
\centering
\includegraphics[width=\textwidth]{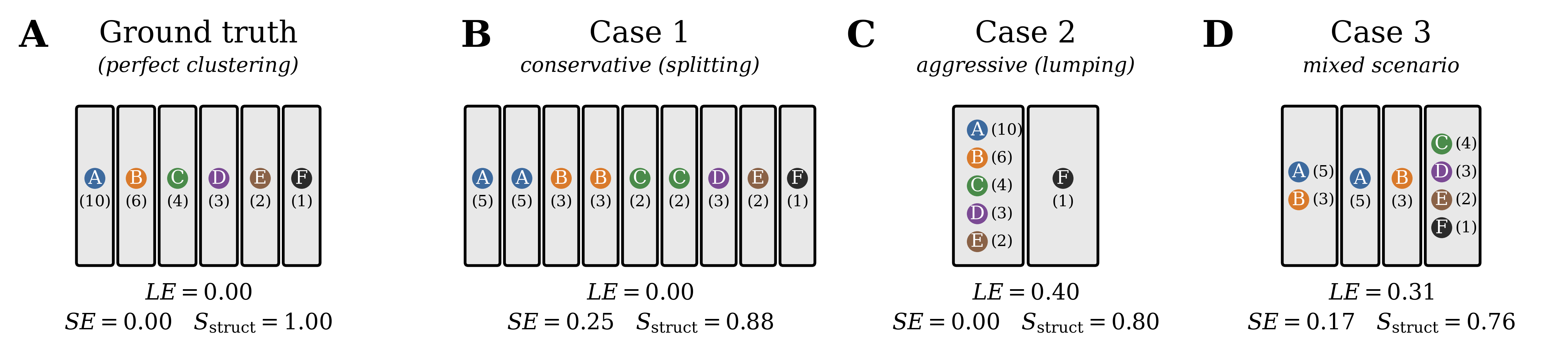}
\caption{Example illustrating the structural interpretation of Lumping Error (LE) and Splitting Error (SE) under different clustering outcomes for six real authors. Numbers in parentheses indicate papers per author (or per fragment, when an author is split).}
\label{fig:toy2}
\end{figure*}

\paragraph{Example: Structural Interpretation of LE and SE.} To provide an intuitive interpretation of the proposed structural metrics, Figure~\ref{fig:toy2} uses the same ground truth as Figure~\ref{fig:toy1} and compares three clustering outcomes: splitting, lumping, and the mixed scenario.

LE quantifies cluster contamination: it increases when publications belonging to different real-world authors are incorrectly merged into the same predicted cluster. Conversely, SE quantifies fragmentation: it increases when publications belonging to the same real-world author are distributed across multiple predicted clusters.

Three representative failure modes are illustrated:
\begin{itemize}
    \item \textbf{Conservative splitting}: predicted clusters remain pure but fragment the same author identity across multiple clusters, resulting in low LE and higher SE.
    \item \textbf{Aggressive lumping}: multiple distinct authors are merged into large clusters, producing high LE and low SE.
    \item \textbf{Mixed scenario}: both contamination and fragmentation coexist, leading to intermediate values for both metrics.
\end{itemize}

Unlike pairwise or B$^3$ metrics, which summarize clustering quality globally, LE and SE explicitly disentangle the two main structural failure modes of author disambiguation systems. This decomposition provides a more interpretable evaluation framework for analysing the behaviour of AND models under different relational conditions.

Table~\ref{tab:error_scenarios} summarizes how pairwise, B$^3$, and structural metrics respond across the three representative failure modes. While pairwise and B$^3$ metrics compress error signals into aggregate F1 scores, LE and SE disentangle the underlying structural patterns.

Conservative configurations achieve perfect precision but incur splitting errors, aggressive configurations maximize recall but introduce lumping errors, and mixed scenarios exhibit moderate trade-offs across all dimensions. This explicit decomposition enables practitioners to diagnose whether a model's errors are driven by over-merging, over-fragmenting, or a combination of both.

\vspace{0.3cm}

\begin{table}[!h]
\centering
\caption{Structural error metrics across disambiguation configurations.}
\label{tab:error_scenarios}
\footnotesize
\setlength{\tabcolsep}{4pt}
\begin{tabular}{l|cccccc}
\toprule
\textbf{Scenario} & $F1_{PW}$ & $B^3_P$ & $B^3_R$ & LE & SE & $S_{\text{struct}}$ \\
\midrule
Ground truth       & 1.00 & 1.00 & 1.00 & 0.00 & 0.00 & 1.00 \\
Case 1 (splitting) & 0.63 & 1.00 & 0.62 & 0.00 & 0.25 & 0.88 \\
Case 2 (lumping)   & 0.38 & 0.29 & 1.00 & 0.40 & 0.00 & 0.80 \\
Case 3 (mixed)     & 0.46 & 0.59 & 0.69 & 0.31 & 0.17 & 0.76 \\
\bottomrule
\end{tabular}
\end{table}

\subsection{Step 3 -- Composite Score and Model Selection}
\label{sec:composite}

Standard metrics may provide conflicting signals when considered independently. To provide a compact, task-oriented criterion for comparing experimental configurations, we define a Composite Score:

\begin{equation}
\label{eq:composite}
\text{Composite Score} = \alpha \cdot F1_{PW} + \beta \cdot F1_{B^3} + \gamma \cdot S_{\text{struct}}
\end{equation}

where $F1_{PW}$ and $F1_{B^3}$ are the harmonic means of the corresponding precision and recall values. We set $\alpha=0.3$, $\beta=0.5$, $\gamma=0.2$, reflecting the following rationale:

\begin{itemize}
    \item $F1_{B^3}$ receives the highest weight ($\beta=0.5$) because B$^3$ satisfies the formal desiderata for clustering evaluation -- homogeneity, completeness, and robustness to cluster size imbalance \cite{amigo2009comparison} -- making it the most reliable single signal in AND, where cluster sizes are highly skewed.
    \item $F1_{PW}$ receives a moderate weight ($\alpha=0.3$) as a global consistency check, capturing pair-level correctness complementary to instance-level B$^3$.
    \item $S_{\text{struct}}$ receives the lowest weight ($\gamma=0.2$) because it plays a diagnostic rather than primary role, penalizing structurally degenerate solutions that may otherwise score reasonably on aggregate F1 metrics.
\end{itemize}

We characterize the robustness of this weighting choice quantitatively in Appendix~\ref{sec:sensitivity}, using the synthetic case study of Appendix~\ref{sec:casestudy} as a testbed.

\paragraph{Model Selection Strategy.} Given the precision-first principle, model selection is not based on a single metric but on the following combination of criteria, in order of priority:

\begin{enumerate}
    \item high $B^3_P$, ensuring cluster purity,
    \item low LE, limiting incorrect merges across authors,
    \item high Composite Score, capturing overall clustering quality.
\end{enumerate}

Configurations that achieve high recall at the expense of increased lumping are not preferred, even if their aggregate F1 scores are competitive. All metrics are reported separately to preserve full interpretability; the Composite Score is used as a compact summary for ranking, not as a replacement for detailed metric analysis. Appendix~\ref{sec:casestudy} gives a worked example of exactly this scenario -- a small Composite Score gap masking a much larger purity gap.

\subsection{Step 4 -- Implementation}

All evaluation metrics are computed within a single unified Python pipeline. The source code is publicly available at \url{https://github.com/fra533/AND-Evaluation-Framework.git}.

The core evaluator operates on two inputs: (i) the predicted clusters $\mathcal{C}_P$ and (ii) the ground-truth clusters $\mathcal{C}_T$. For datasets where not all ground-truth papers are retrievable, the ground-truth clusters are filtered prior to evaluation to retain only papers present in the predicted output, ensuring that recall is not artificially penalised by missing data (see Section~\ref{sec:discussion}).

\paragraph{Indexing.} For each author name, two lookup maps are constructed:

\begin{algorithm}[H]
\caption{Index Construction}
\label{alg:indexing}
\begin{algorithmic}[1]
\State \textbf{Input:} Papers $x$ in both predicted and ground-truth datasets
\State \textbf{Output:} Two lookup maps
\State $p\_index[x] \gets$ predicted cluster index for paper $x$
\State $t\_index[x] \gets$ ground-truth cluster index for paper $x$
\end{algorithmic}
\end{algorithm}

\paragraph{Pairwise metrics.} TP and FP are computed by enumerating all pairs within each predicted cluster and checking ground-truth co-membership; FN is computed symmetrically:

\begin{algorithm}[H]
\caption{Pairwise Metrics Computation}
\label{alg:pairwise}
\begin{algorithmic}[1]
\State \textbf{Input:} Predicted clusters $\mathcal{C}_P$, Ground-truth clusters $\mathcal{C}_T$
\State \textbf{Output:} $TP$, $FP$, $FN$ counts
\State $TP \gets 0$, $FP \gets 0$, $FN \gets 0$
\For{each cluster $C$ in $\mathcal{C}_P$}
    \For{each pair $(a, b)$ in $C$}
        \If{$t\_index[a] = t\_index[b]$}
            \State $TP \gets TP + 1$
        \Else
            \State $FP \gets FP + 1$
        \EndIf
    \EndFor
\EndFor
\For{each cluster $T$ in $\mathcal{C}_T$}
    \For{each pair $(a, b)$ in $T$}
        \If{$p\_index[a] \neq p\_index[b]$}
            \State $FN \gets FN + 1$
        \EndIf
    \EndFor
\EndFor
\State \textbf{Return:} $TP$, $FP$, $FN$
\end{algorithmic}
\end{algorithm}
\paragraph{B$^3$ metrics.} Computed in a single pass over all papers in \texttt{p\_index} $\cap$ \texttt{t\_index}:

\begin{algorithm}[H]
\caption{B$^3$ Metrics Computation}
\label{alg:b3}
\begin{algorithmic}[1]
\State \textbf{Input:} Papers in $p\_index \cap t\_index$, $N$ (total papers)
\State \textbf{Output:} $B^3_P$ (precision), $B^3_R$ (recall)
\State $B^3_P \gets 0$, $B^3_R \gets 0$
\For{each paper $x$ in $p\_index \cap t\_index$}
    \State $inter \gets |P(x) \cap T(x)|$
    \State $B^3_P \gets B^3_P + inter / |P(x)|$
    \State $B^3_R \gets B^3_R + inter / |T(x)|$
\EndFor
\State $B^3_P \gets B^3_P / N$
\State $B^3_R \gets B^3_R / N$
\State \textbf{Return:} $B^3_P$, $B^3_R$
\end{algorithmic}
\end{algorithm}

\paragraph{Structural metrics.} LE and SE are computed at the cluster level for each author name, then averaged across names (name-level macro-average):

\begin{algorithm}[H]
\caption{Structural Metrics Computation (LE, SE, $S_{\text{struct}}$)}
\label{alg:structural}
\begin{algorithmic}[1]
\State \textbf{Input:} Predicted clusters $\mathcal{C}_P$, Ground-truth clusters $\mathcal{C}_T$
\State \textbf{Output:} $LE$, $SE$, $S_{\text{struct}}$
\State $LE \gets 0$, $SE \gets 0$
\For{each cluster $C$ in $\mathcal{C}_P$}
    \State $g \gets$ number of distinct $t\_index$ values in $C$
    \State $LE \gets LE + \max(0, g-1) / g$
\EndFor
\State $LE \gets LE / |\mathcal{C}_P|$
\For{each cluster $T$ in $\mathcal{C}_T$}
    \State $p \gets$ number of distinct $p\_index$ values in $T$
    \State $SE \gets SE + \max(0, p-1) / p$
\EndFor
\State $SE \gets SE / |\mathcal{C}_T|$
\State $S_{\text{struct}} \gets 1 - (LE + SE) / 2$
\State \textbf{Return:} $LE$, $SE$, $S_{\text{struct}}$
\end{algorithmic}
\end{algorithm}

Averaging first within each author name and then across names ensures that authors associated with many clusters do not disproportionately dominate the structural error signal. Appendix~\ref{sec:casestudy} quantifies exactly how much this choice matters relative to a micro-average.

\paragraph{Composite Score.} Computed as the weighted sum of pairwise F1, instance-level $B^3$ F1, and the structural score $S_{\text{struct}}$:

\begin{algorithm}[H]
\caption{Composite Score Computation}
\label{alg:composite}
\begin{algorithmic}[1]
\State \textbf{Input:} $F1_{PW}$, $F1_{B^3}$, $S_{\text{struct}}$
\State \textbf{Output:} Composite Score
\State \textbf{Weights:} $\alpha = 0.3$, $\beta = 0.5$, $\gamma = 0.2$
\State $\text{Score} \gets \alpha \cdot F1_{PW} + \beta \cdot F1_{B^3} + \gamma \cdot S_{\text{struct}}$
\State \textbf{Return:} Score
\end{algorithmic}
\end{algorithm}

\section{Discussion}
\label{sec:discussion}
The illustrative example of Table~\ref{tab:error_scenarios} shows what conventional metrics usually fail to characterize in AND system behavior. Cases 2 and 3 obtain similar pairwise F1 scores (0.38 and 0.46) and similar $S_{\text{struct}}$ (0.80 and 0.76), yet exhibit different structural patterns: Case 2 fails only through aggressive lumping ($LE = 0.40$, $SE = 0$), while Case 3 combines contamination and fragmentation ($LE = 0.31$, $SE = 0.17$). Aggregate scores collapse these qualitatively different failure modes into similar values; only the separate inspection of LE and SE reveals them.
 
The controlled case study in Appendix~\ref{sec:casestudy} (Tables~\ref{tab:casestudy-data}--\ref{tab:casestudy-results}) shows the same effect at the level of model selection: Configs A and C differ by only 0.023 in Composite Score, yet C loses 17 points of $B^3_P$ and introduces non-zero lumping ($LE=0.083$). The ordered Model Selection Strategy rejects C, whereas the Composite Score alone would treat the two as roughly comparable.
 
By decoupling LE and SE at the cluster level, our framework provides structural diagnosis. LE measures whether a model is too aggressive in merging; SE measures whether it is too conservative in fragmenting. This distinction matters operationally: lumping contaminates author profiles and propagates downstream, while splitting is often recoverable through complementary signals (citations, co-authorship networks, ORCID disambiguation). The Model Selection Strategy (Section~\ref{sec:composite}) leverages this asymmetry by prioritizing B$^3_P$ and LE inspection before the Composite Score, making the cost trade-off explicit.

Sensitivity analysis (Appendix~\ref{sec:sensitivity}, Tables~\ref{tab:sweep-gamma}--\ref{tab:sweep-alphabeta}) shows that the adopted weights for the Composite Score ($\alpha=0.3$, $\beta=0.5$, $\gamma=0.2$) lie within a stable region of the weight space: the ranking holds for $\gamma \in [0, 0.5]$, inverts only at $\gamma \geq 0.6$, and is stable across the full $\alpha{:}\beta$ range.
 
The framework encodes a precision-first stance, prioritizing cluster purity over completeness. This aligns with production AND systems where false associations are costly to correct post-hoc. However, this design choice is deliberate, not universal: applications where completeness is paramount (retrospective archival work, historical attribution) can modify the Model Selection Strategy hierarchy to inspect SE before LE, or adjust the Composite Score weights to elevate $\gamma$ and increase structural metric influence.
 
The framework makes this trade-off transparent and actionable, allowing practitioners to calibrate the balance for their specific operational context.
 
Four limitations are important to acknowledge. First, partial ground-truth coverage (common in open bibliographic datasets) requires filtering ground-truth clusters to match predicted output, making results dataset-specific. Second, pairwise metric computation scales as $O(n^2)$ per cluster, creating a bottleneck for author names with very large paper sets. Third, macro-averaging of LE/SE treats all author names equally, preventing high-ambiguity names from dominating the aggregate signal but reducing sensitivity to systematic failures on highly complex names (quantified in Table~\ref{tab:macro-micro}). Fourth, the symmetric treatment of LE and SE in $S_{\text{struct}}$ does not inherently carry the precision-first asymmetry, necessitating the ordered-priority Model Selection Strategy.

\section{Conclusions}

We present a structured, precision-driven evaluation framework for AND that bridges a critical gap in current evaluation practice. It decomposes clustering quality into three complementary perspectives (pairwise consistency, B$^3$ instance-level metrics, and cluster-level structural error measures) integrated into a single Composite Score for model selection. 

The framework has been validated on a controlled synthetic case study, tested for robustness to weight choices, it enables practitioners to diagnose whether model errors stem from over-merging distinct authors or over-fragmenting single author profiles.
This structural interpretation is operationally valuable in production settings where these error types carry asymmetric downstream costs. The framework is implemented as an open-source Python pipeline designed for practical interoperability across AND benchmarks.

We believe this framework advances the AND evaluation landscape toward more principled, interpretable, and diagnostically rich assessment practices.

\section{Acknowledgments}

\section{Declaration of Use of Generative AI}
No generative AI systems were used to generate or analyse data, or draft the scientific content of this manuscript. AI-assisted tools were used only for spell-checking and minor grammar suggestions on the English text. Generative AI–based coding assistants were used during (the development to support code generation and, especially, review); all AI-suggested code was reviewed, tested, and validated by the authors.

%



\appendix
 
\section{Synthetic Case Study and Sensitivity Analysis}
\label{app:validation}
 
This appendix provides the controlled case study referenced in Section~\ref{sec:composite} and the sensitivity analysis of the Composite Score weights discussed in Section~\ref{sec:discussion}.
 
\subsection{Synthetic case study}
\label{sec:casestudy}
 
To ground the claims of Sections~\ref{sec:precision-first}--\ref{sec:composite} in a controlled, reproducible setting, we construct a small synthetic dataset of five ambiguous name strings with heterogeneous ambiguity levels (Table~\ref{tab:casestudy-data}), and evaluate three synthetic clustering configurations against it -- chosen to span the precision--recall spectrum this framework is designed to characterize.
 
\begin{table}[h]
\centering
\caption{Synthetic dataset composition.}
\label{tab:casestudy-data}
\small
\setlength{\tabcolsep}{8pt}
\renewcommand{\arraystretch}{1.1}
\begin{tabular}{@{}lcc@{}}
\toprule
\textbf{Name} & \textbf{Real authors ($|\mathcal{C}_T|$)} & \textbf{Papers} \\
\midrule
N1 & 4 & 12 \\
N2 & 3 & 9 \\
N3 & 2 & 6 \\
N4 & 2 & 5 \\
N5 & 1 & 3 \\
\midrule
\textbf{Total} & \textbf{12} & \textbf{35} \\
\bottomrule
\end{tabular}
\end{table}
 
We define three configurations:
\begin{itemize}
    \item \textbf{Config A (Conservative):} Every predicted cluster is pure ($g(C)=1$); recall loss comes only from occasionally splitting a real author's output into two clusters (N1: two authors split; N2: one author split; N3: one author split; N4--N5: perfectly recovered).
    \item \textbf{Config B (Aggressive/Uniform Lumping):} Predicted clusters merge real authors across every ambiguous name (N1: two 2-author merges; N2: all three authors merged into one cluster; N3--N4: the two real authors merged into one cluster each; N5 is unaffected, being unambiguous).
    \item \textbf{Config C (Targeted/Moderate Lumping):} Lumping is confined to the two most ambiguous names (N1, N2 -- one merge of two real authors each); N3--N5 are recovered perfectly, with no splitting.
\end{itemize}
 
\begin{table*}[t]
\centering
\caption{Results across configurations (name-level macro-average for LE/SE).}
\label{tab:casestudy-results}
\resizebox{0.92\textwidth}{!}{%
\renewcommand{\arraystretch}{1.15}
\begin{tabular}{@{}lcccccccc@{}}
\toprule
\textbf{Config.} & \textbf{$B^3_P$} & \textbf{$B^3_R$} & \textbf{$F1_{B^3}$} & \textbf{$F1_{PW}$} & \textbf{LE} & \textbf{SE} & \textbf{$S_{\text{struct}}$} & \textbf{Composite} \\
\midrule
A (Conservative)       & 1.0000 & 0.8476 & 0.9175 & 0.8667 & 0.0000 & 0.1333 & 0.9333 & \textbf{0.9054} \\
C (Targeted lumping)   & 0.8286 & 1.0000 & 0.9062 & 0.7907 & 0.0833 & 0.0000 & 0.9583 & 0.8820 \\
B (Aggressive lumping) & 0.5029 & 1.0000 & 0.6692 & 0.5312 & 0.4333 & 0.0000 & 0.7833 & 0.6506 \\
\bottomrule
\end{tabular}%
}
\end{table*}
 
Ranked by Composite Score alone (Table~\ref{tab:casestudy-results}), A and C are close -- a gap of only 0.023 (0.9054 vs.\ 0.8820), which in isolation could read as ``roughly comparable.'' But $B^3_P$ drops by 17 points (1.00 $\to$ 0.83) and LE rises from a perfectly pure 0.000 to 0.083 -- a non-trivial contamination rate concentrated on the two hardest names. This is precisely why the Model Selection Strategy of Section~\ref{sec:composite} inspects $B^3_P$ and LE \emph{before} the Composite Score: a small Composite gap can mask a purity difference that matters far more under the lumping/splitting cost asymmetry motivating this framework. Config B, by contrast, is dominated on every metric and is rejected unambiguously by all three criteria -- it is Config C, not B, that demonstrates why the ordered multi-criteria check is necessary rather than the Composite Score in isolation.
 
\begin{table}[H]
\centering
\caption{Macro- vs.\ micro-averaging of LE and SE.}
\label{tab:macro-micro}
\small
\setlength{\tabcolsep}{5pt}
\renewcommand{\arraystretch}{1.1}
\begin{tabular}{@{}lcccc@{}}
\toprule
\textbf{Config.} & \multicolumn{2}{c}{\textbf{LE}} & \multicolumn{2}{c}{\textbf{SE}} \\
\cmidrule(lr){2-3} \cmidrule(lr){4-5}
& \textbf{Macro} & \textbf{Micro} & \textbf{Macro} & \textbf{Micro} \\
\midrule
A & 0.000 & 0.000 & 0.133 & 0.167 \\
C & 0.083 & 0.100 & 0.000 & 0.000 \\
B & 0.433 & 0.444 & 0.000 & 0.000 \\
\bottomrule
\end{tabular}
\end{table}
 
Micro-averaging (Table~\ref{tab:macro-micro}) pools all clusters across names before computing LE/SE, so names with more ground-truth clusters (N1: 4, N2: 3) contribute more terms and pull the aggregate toward their own error rate. The effect is visible but modest here (e.g., Config A's SE moves from 0.133 to 0.167 under micro-averaging) because our synthetic ambiguity spread is mild; in real name-ambiguity distributions -- which are typically heavy-tailed (where max real-authors-per-name reaches 96 against a mean of 13--20) -- this divergence would be substantially larger. We report both here to make the direction and magnitude of the effect concrete rather than only asserted.
 
\subsection{Sensitivity of the Composite Score}
\label{sec:sensitivity}
 
The weights $\alpha=0.3$, $\beta=0.5$, $\gamma=0.2$ reflect our precision-first stance and the relative reliability of B$^3$ as a clustering signal. We quantify their robustness using the three configurations of Section~\ref{sec:casestudy}, first varying $\gamma$ (Table~\ref{tab:sweep-gamma}) and then the $\alpha{:}\beta$ split at fixed $\gamma$ (Table~\ref{tab:sweep-alphabeta}).
 
\begin{table*}[t]
\centering
\caption{Composite Score under varying structural weight $\gamma$. The ratio $\alpha:\beta$ is fixed at $0.3:0.5$ and rescaled to sum to $1-\gamma$.}
\label{tab:sweep-gamma}
\small
\setlength{\tabcolsep}{3.5pt}
\renewcommand{\arraystretch}{1.1}
\begin{tabular}{@{}ccccccc@{}}
\toprule
$\gamma$ & $\alpha$ & $\beta$ & \textbf{A} & \textbf{C} & \textbf{B} & \textbf{Ranking} \\
\midrule
0.0 & 0.375 & 0.625 & 0.898 & 0.863 & 0.617 & A $>$ C $>$ B \\
0.1 & 0.337 & 0.562 & 0.902 & 0.872 & 0.634 & A $>$ C $>$ B \\
\textbf{0.2} & \textbf{0.300} & \textbf{0.500} & \textbf{0.905} & \textbf{0.882} & \textbf{0.651} & \textbf{A $>$ C $>$ B} \\
0.3 & 0.262 & 0.438 & 0.909 & 0.892 & 0.667 & A $>$ C $>$ B \\
0.4 & 0.225 & 0.375 & 0.912 & 0.901 & 0.684 & A $>$ C $>$ B \\
0.5 & 0.187 & 0.312 & 0.916 & 0.911 & 0.700 & A $>$ C $>$ B \\
0.6 & 0.150 & 0.250 & 0.919 & \textbf{0.920} & 0.717 & \textbf{C $>$ A $>$ B} \\
\bottomrule
\end{tabular}
\end{table*}
 
As shown in Table~\ref{tab:sweep-gamma}, the ranking A $\succ$ C $\succ$ B is stable across the region around our adopted weights ($\gamma \in [0, 0.5]$), but inverts to C $\succ$ A at $\gamma \geq 0.6$. The mechanism is the one anticipated in Section~\ref{sec:structural}: $S_{\text{struct}}$ weights LE and SE symmetrically, so Config C ($LE{=}0.083$, $SE{=}0$, $S_{\text{struct}}{=}0.958$) scores higher on $S_{\text{struct}}$ than Config A ($LE{=}0$, $SE{=}0.133$, $S_{\text{struct}}{=}0.933$) even though A is the only one of the two with zero lumping. As $\gamma$ grows, this symmetric component increasingly dominates the Composite Score and eventually overrides the asymmetric preference encoded by our precision-first stance. At our adopted $\gamma=0.2$, the margin to this inversion point is comfortable, and the second sweep over the $\alpha{:}\beta$ split (Table~\ref{tab:sweep-alphabeta}) shows the ranking is stable across the full range at fixed $\gamma$ -- instability is specific to $\gamma$, not to the relative weighting of the two aggregate F1 components.
 
\begin{table*}[h]
\centering
\caption{Composite Score under varying $\alpha:\beta$ split, with fixed $\gamma=0.2$.}
\label{tab:sweep-alphabeta}
\small
\setlength{\tabcolsep}{6pt}
\renewcommand{\arraystretch}{1.1}
\begin{tabular}{@{}ccccc@{}}
\toprule
$\alpha$ & $\beta$ & \textbf{Comp. A} & \textbf{Comp. C} & \textbf{Comp. B} \\
\midrule
0.0 & 0.8 & 0.921 & 0.917 & 0.692 \\
0.2 & 0.6 & 0.911 & 0.894 & 0.664 \\
\textbf{0.3} & \textbf{0.5} & \textbf{0.905} & \textbf{0.882} & \textbf{0.651} \\
0.5 & 0.3 & 0.895 & 0.859 & 0.623 \\
0.8 & 0.0 & 0.880 & 0.824 & 0.582 \\
\bottomrule
\end{tabular}
\end{table*}
 
We surface this as a limitation rather than a defect: $S_{\text{struct}}$'s symmetric treatment of LE and SE is a deliberate simplification for interpretability (Section~\ref{sec:structural}), and $\gamma=0.2$ keeps its influence on the final ranking modest by design. Practitioners adopting a substantially higher $\gamma$ should be aware that $S_{\text{struct}}$ in its current form does not itself encode the lumping/splitting cost asymmetry, and should inspect LE directly -- per the Model Selection Strategy of Section~\ref{sec:composite} -- rather than relying on a high-$\gamma$ Composite Score alone.
 
\end{document}